\documentclass{aa}
\newcommand{\micron}{\,\mu{\rm m}}
\input epsf
\begin{document}

   \thesaurus{	10.19.3;  
		13.09.1   
	}
   \title{Evidence for a two-armed spiral in the Milky Way}

   \author{R. Drimmel}

   \institute{Osservatorio Astronomico di Torino, 10025 Pino Torinese, Italy\\
              email: drimmel@to.astro.it
             }

   \date{Received ; accepted }

   \maketitle

   \begin{abstract}

Emission profiles of the Galactic plane in K and at $240\micron$ are
presented, and features associated with the tangents of the spiral
arms are identified. In the K band, which traces stellar emission
and suffers little from absorbtion, features
associated with the arm tangents indicate that
a two-armed logarithmic spiral dominates the nonaxisymmetric
structure of the Milky Way.
In contrast, the $240\micron$ emission from dust entrained in the
interstellar gas is consistent with a four-armed model, 
in concordance with radio data and optical spiral tracers.
This suggests that the non-axisymmetric mass perturbation
responsible for the four-armed spiral structure in the gas 
has a two rather than four-armed structure.

\keywords{Galaxy: structure; Infrared: galaxies}

   \end{abstract}

%

\section{Introduction}

A nearly ubiquitous and prominent feature of disk galaxies is the
presence of spiral arms. Unfortunately the spiral arm structure of
our own Galaxy is much less obvious due to our obscured view
from within its disk.
Luminous spiral arm tracers, especially HII regions, have
been used to effectively map the main spiral arms of our Galaxy 
(Georgelin and Georgelin \cite{GG76}),
with the help of kinematic distances. Similar mappings have
been made with components of the gaseous medium, such as HI
and CO, and more recently the direction of the galactic magnetic field,
which is presumably aligned with the arms, has been inferred from
studies of pulsar radio observations.
From such studies it is usually inferred that the
Milky Way has four spiral arms with a pitch angle of approximately
$12 \degr$ (see Vallee \cite{val95} for a recent review).
Apart from such spatial mappings, three of the spiral arms directly evidence
themselves in Galactic plane emission profiles
as features in the directions of the arm tangents
where the column flux density of the associated material is greatest.

If the spiral structure is produced by a non-axisymmetric
component in the stellar distribution, where the
majority of the disk mass resides, then the observed gas and optical
spiral tracers make up a small fraction of the total mass actually
associated with the spiral arms.
To directly observe the mass associated with spiral structure one must observe
the old stellar population, which is best traced by K band imagery
(Rix and Zaritsky \cite{Rix95}). 
Alternatively, the spiral structure could be produced by
self-propagating star formation in a differentially rotating
medium, which has been shown with simulations to produce
spiral arms (Seiden et al. \cite{Seiden79}, 
Jungwiert and Palous \cite{Jung94}). In this case the spiral pattern in
K would be produced solely by young stars, which can contribute a significant
fraction of the K band light in star forming regions (Rhoads \cite{Rho98}),
and will have a form consistent with the optical spiral tracers.

In this letter the K band and 240$\micron$ emission profiles
of the Galactic plane are recovered from COBE/DIRBE data,
and considered with regards to the spiral structure
of the Milky Way. In the following section features
associated with spiral arm tangents are identified, and in
Sect. 3 the number and position of the tangents are
interpreted with a simple logarithmic spiral model.


\section{Galactic plane emission profiles}

Galactic plane (GP) emission profiles
in the K and 240$\micron$ bands were constructed from the
'Zodi-Subtracted Mission Average (ZSMA)'
sky maps produced from the COBE/DIRBE data (Kelsall at al. \cite{dirbe2}).
From the high resolution intensity maps (pixel width $\approx .35 \degr$)
pixels within three degrees of the GP where integrated over galactic
latitude after reprojecting the data into a Galactic
Mollweide projection, using the UIDL analysis package developed by
NASA's Goddard Space Flight Center. The resulting emission profiles
(integrated intensity) are shown for galactic longitudes
$|l| < 90\degr$ in Figure \ref{gpprof}. The K band emission profiles
for $|b| > 3 \degr$
show no evidence of spiral arm features, with the exception of longitudes
$-60 > l > -90 \degr$.

Emission within $15\degr$ of the Galactic center (GC) is dominated
by light from the bulge of the Galaxy and absorbtion from its
associated dust lanes, while
the very broad and prominent feature at approximately $80 \degr$
in the 240$\micron$ band is the Orion arm.
This arm is a local and relatively minor structure in the
Galactic disk and not a major spiral arm
(Georgelin \& Georgelin \cite{GG76}); it is prominent
only by virtue of our proximity and will
not be considered further. I only mention here
that it's counterpart, centered at $l \approx -100\degr$,
is just outside the right bounds of the figure. Neither do I attempt
to identify features associated with this local arm or the Galactic bulge.

\begin{figure*}
\epsfysize=14cm
\epsffile{cd132.f1}
\caption{Galactic plane ($|b| < 3 \degr$) emission profiles for K band
(squares; left vertical scale) and at 240$\micron$ (triangles;
right vertical scale) on a logarithmic scale.
Diamonds at $l < -60 \degr$ show K band emission integrated over $|b| < 9 \degr$.
Peaks of major features are identified with vertical lines: broad,
prominent features in K (bold solid lines); strong, relatively narrow
features in K (dash-dotted lines); broad features at $240\micron$
(dashed lines).
These features are grouped and labeled according to proximity in
longitude. The horizontal error bars above the K band profile show the
tangent locations of a two-arm logarithmic spiral assuming the
uncertainties shown for the S$_1$ and S$_2$ tangents in K,
while those below the
$240 \micron$ profile show the tangent locations of a four-arm spiral,
assuming the indicated uncertainties for the S$_2$ and C$_2$ tangent
locations at $240 \micron$.
The emission within $15\degr$ of the Galactic center is dominated by the
bulge, while the $240 \micron$ feature at $80\degr$ is the (local) Orion
arm.
}
\label{gpprof}
\end{figure*}

Features associated with the tangents to the major spiral arms
are expected appear in the galactic longitude range
$l > -80 \degr$, $l < 60 \degr$.
In this longitude range three types of features are identified and their peaks
indicated with vertical lines in Fig. \ref{gpprof}:
broad ($\Delta l \approx 10\degr$), prominent features
in the K band; strong narrow features in K; and peak emission of
broad features at $240\micron$.
At negative galactic longitudes the spiral structure of the Galaxy shows
itself clearly in K as two broad features at $\approx -20$ and $-50\degr$, 
and two narrow features at $\approx -70 \degr$,
while the $240\micron$ emission has a distinct sawtooth profile.
At positive longitudes the spiral arms are less
evident, with a single prominent feature at 
$\approx 30 \degr$ double peaked in both the K and 240$\micron$ bands.
In K the appearance of a double peak can be attributed to absorbtion, as the
240$\micron$ peak at $31\degr$ exactly corresponds
to the ``valley'' between the peaks seen in K.

The identified features can be grouped by proximity in longitude
in the following manner:
at negative galactic longitudes there are three groups, labeled
``T'', ``S$_2$'' and ``C$_2$'', while at positive longitudes
there are the groups ``S$_1$'' and ``C$_1$''. Though the C$_1$
group consists of only a single feature at $240\micron$ it is labeled to
facilitate the following discussion.
I consider the narrow feature at $42\degr$
not to be sufficiently close to other features to be part of a group
and leave it unlabeled. With the exception of C$_1$,
each group consists of both 240$\micron$ and K band features and span
approximately $10\degr$ or more in longitude. All other unlabeled
features in the K band, in the considered longitude range, have
a relative peak to local background signal of less than one fourth.

Emission profiles at other wavebands have not been presented here.
In the far-infrared COBE also observed at 100 and 140$\micron$,
which present very similar GP profiles as the 240$\micron$ data.
In the near-infrared J, L and M band observations were made, but these
are not as cleanly interpreted. The M band is contaminated with
Zodiacal emission, while the J band suffers significantly
from the effects of absorbtion in the GP. The L band is
in principle more transparent than the K band, however, it
is not as reliable a tracer of the stellar mass in the GP as it
suffers from molecular emission from polycyclic aromatic hydrocarbons
(Dwek  et al. \cite{dwek97}).
For instance, in L the S$_1$ feature appears as a single peak,
suggesting absorbtion at play in K, while 
the S$_2$ feature appears double peaked, the second
peak appearing at precisely the longitude of the 240$\micron$
peak; this is a clear indication of L band emission associated with the
dust.


\section{Interpretation}

Each group of features in the GP profiles can be identified
with a spiral arm tangent. Three of these groups possess
broad K band features: S$_1$ and S$_2$ are tangents
of the Scutum arm, while the third, T, is the Three-kiloparsec arm.
The remaining groups, C$_1$ and C$_2$, can be identified as tangents
to the Sagittarius-Carina (Sag-Car) arm, though there are important
differences in their characteristics with respect to
the others in the K band. First, the feature at C$_2$ has a distinctly
different character than the broad features mentioned above, being composed of
two narrow features. (The profile for $|b| < 9 \degr$ does not
blend or enhance these features above the background.)
These narrow features could be combined and
considered as a single feature, but its apparent width would still
be considerably less than the broad tangent features if
removed to similar heliocentric distances
(from .4 -- .5 to .8 -- 1.0 $R_\odot$).
More serious is the absence of a significant feature in K at
$\approx 50\degr$, also noted in other infrared studies
(see Table 1 of Ortiz and Lepine \cite{Ortiz93}).
If the Sag-Car arm was of similar amplitude as the Scutum
arm it would produce a tangent feature at $\approx 50\degr$
of similar amplitude as the S$_2$ feature,
since spiral arm tangents at $\pm 50\degr$ would
have approximately the same galactocentric distance.

From the galactic longitudes of two tangents the
parameters of a logarithmic
spiral described by $r=A e^{-a \phi}$ can be determined,
where $(r,\phi)$ are galactocentric cylindrical coordinates.
Geometrical considerations give the tangent of the
pitch angle $p$, without approximation, as
\begin{equation}
\label{tani}
\tan p = a = \frac {\ln \left( \frac{\sin | l_- |}{\sin l_+} \right)}
                {\pi - (l_+ - l_-)},
\end{equation}
where $l_-$ and $l_+$ are the longitudes of the tangents at
negative and positive longitudes respectively.
Using the K band S$_1$ and S$_2$ features at $(l_+,l_-) = (27, -53) \degr$,
one arrives at $p = 17.9\degr$. However, if the double peak of the S$_1$
feature is due to strong absorbtion, then its true tangent may be
at $l_+ = 30\degr$ longitude. In this case $p = 15.5\degr$.
In addition, a bias is
introduced in the apparent tangent direction as the
integrated emission comes from a curved structure with finite
width. The peak emission is thus displaced toward the GC, the direction of
curvature for the arm. A two degree uncertainty has
been estimated for the S$_2$ tangent longitude due to this bias and
possible absorbtion effects.

The galactic longitude of other arm tangents can be found by using the
pitch angle as found above and numerically
solving for $\theta$ in the equation
\begin{equation}
e^{-a \theta} \sin \theta = e^{a ( \Delta \phi - \theta_0 )} \sin \theta_0,
\label{tang}
\end{equation}
the longitude angle $\theta_0$ being a given tangent. Using the position
of the S$_2$ tangent for $\theta_0$
and $\Delta \phi = - \pi$, the tangent for the counter spiral of
the Scutum arm is found to be at
$-20.3 < \theta < -22.8\degr$, corresponding well to the observed
longitude of the Three-kiloparsec arm tangent. A second tangent from this arm
is not expected at positive galactic longitudes, as it would pass
beyond the Solar Circle.

At $240 \micron$ a four-armed model can be applied.
The features with the best defined peaks
at this wavelength are S$_2$ and C$_2$, tangents of adjacent arms.
Setting $\Delta \phi = - \pi/2$ in Eq. \ref{tang} one can retrieve
a formula similar to Eq. \ref{tani} for $a$, and use the same approach
as above to derive the other arm tangents. Assuming a two degree
uncertainty in the positions of the S$_2$ and C$_2$ peaks,
the pitch angle is found to be between 11.5 and 14.2$\degr$ ,
consistent with previous determinations. The
locations of the predicted arm tangents are shown in Fig. \ref{gpprof}.


\section{Discussion}

A comparison of the spiral arm features in the K band and $240 \micron$
emission profiles show important differences. 
The broad, prominent features in K are consistent with a two-armed spiral
model, and show significant offsets from associated $240 \micron$
features, while the $240 \micron$ emission is consistent with the
traditional four-arm model. Specifically,
there are no significant features at approximately 50 and $-35\degr$ in K
as would be expected from a four-arm model, while the C$_2$ (Sag-Car)
feature shows distinct differences in relative size and form from 
the three broad features at S$_1$, S$_2$ and T.

\begin{figure}[ht]
\epsfysize=8cm
\epsffile{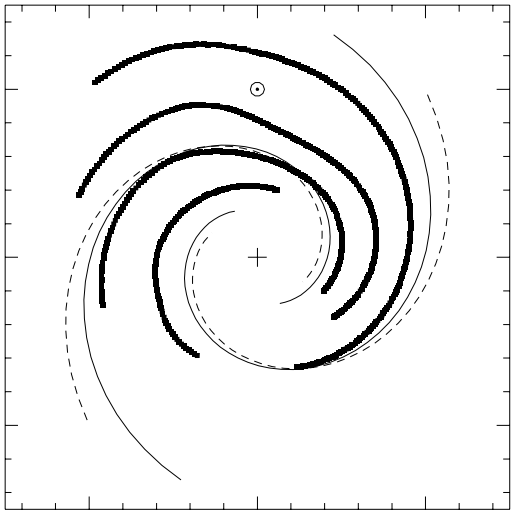}
\caption{The two-arm spiral seen in K
compared with those seen in the optical (bold lines), as traced by
observed HII regions. Solid and dashed lines show the spirals
with the minimum ($15.5\degr$) and maximum (19$\degr$) pitch angles
permitted by the positions of the observed
K band tangent directions and their uncertainties.
The HII arms as illustrated are sparse in the quadrant opposite the Sun ($\odot$)
due to lack of data.
}
\label{sparm}
\end{figure}

The disparity between the $240 \micron$ and K band emission is most
cleanly interpreted as indicating that the diffuse stellar
emission associated with the arms is not primarily from young stars,
but from a nonaxisymmetric density variation in the old stellar
population dominated by a two-arm structure.
This interpretation is consistent with the
distribution of bright near-infrared point sources in the GP, which
can be modeled with a traditional four-armed spiral component
restricted to the distribution of the youngest stellar populations
(Wainescoat et al. \cite{Sky92}; Ortiz and Lepine \cite{Ortiz93}; 
Hammersley et al. \cite{Sky99}); the brightest (youngest) stars
show the same structure as the optical spiral tracers,
in contrast to the fainter (older) stars as
evidenced by the diffuse near-infrared emission.
 Fig. \ref{sparm} compares
the two-arm spiral seen in the diffuse K band emission with
those mapped by HII regions (Taylor and Cordes \cite{TC93}).

The spiral structure suggested by the K band emission profile for
our Galaxy is consistent with near-infrared observations of
external galaxies, in that the arms are seen to be broad,
well described as logarithmic, and more open than their counterparts
at optical wavelengths (Rix and Zaritsky \cite{Rix95}; 
Grosbol and Patsis \cite{Gros98}). Also worth
noting is that two-armed modes are most common in K band observations
of external disk galaxies, regardless of their structure
at visible bands, which often can show little correlation
with the spirals in the near-infrared (Seigar and James \cite{Seig98b}).
The contrast between the optical and infrared spiral structure can be
understood by remembering that optical spiral tracers
are products of star formation in the gaseous arms.
The optical spiral tracers, with the dust emission and bright
infrared point sources, thus trace
the response of the gas to an underlying nonaxisymmetric
stellar (mass) distribution;
what is unique about the diffuse K band emission is that it principally
traces this mass distribution.

The differences in the nonaxisymmetric structure in the gas and stars
noted by the above extragalactic studies and in this contribution
indicate that the hydrodynamical response to a given potential
is not necessarily a simple one.
It has been noted in hydrodynamical simulations that the gaseous
response to a barred potential can have an extended four-armed spiral structure
(Englmaier and Gerhard \cite{Engl99}; Fux \cite{Fux99}), 
while Patsis et\,al (\cite{Patsis97}) have reproduced
extensive interarm features with two-armed potentials.
These simulations suggest an explanation for the relative
weakness of the Sag-Car arm: it is an interarm or secondary
arm structure, possibly bifurcating from one of the primary arms of the
Galaxy. However, to date dynamical models of the Milky Way's stellar disk
that include spiral arms have used the tangents seen in optical
spiral tracers {\em as constraints} (Amaral and Lepine \cite{AL97}).
The disparity between the diffuse K band and optical/gas spirals
suggest that near-infrared observations
will provide crucial information for reconstructing the
dynamical and hydrodynamical processes responsible for
producing spiral structure.

\begin{acknowledgements}
For useful discussions and encouragement the author would like to acknowledge
David Spergel, Ortwin Gerhard and Mario Lattanzi.
\end{acknowledgements}


%
%
%
%
%
%
%
%
%
%

\end{document}